\documentclass[pra,aps,nofootinbib,showpacs,floatfix,superscriptaddress,amsmath,twocolumn]{revtex4-1}

\usepackage{graphicx}
\usepackage{amsmath}
\usepackage{amssymb}
\usepackage{color}
\usepackage{bm}

\newcommand{\unit}{1\!\!1}
\newcommand {\fabs}[1] {\left| #1 \right|}

\newcommand {\fabsq}[1] {\left| #1 \right|^2}

\newcommand{\ket}[1]{\ensuremath{|#1\rangle}}
\newcommand{\bra}[1]{\langle#1|}

\newcommand{\ketbra}[2]{|#1\rangle\langle#2|}
\newcommand{\braXket}[3]{\langle#1|#2|#3\rangle}

\newcommand{\cF}{{\cal{F}}}

\newcommand{\supplement}{the appendix }
\newcommand{\Hamil}{\mathcal{H}}

\graphicspath{{./images/}}

\begin{document}
\title{Quantum Control via Enhanced Shortcuts to Adiabaticity}
\author{C. Whitty}
\email{c.whitty@umail.ucc.ie}
\affiliation{Department of Physics, University College Cork, Cork, Ireland}

\author{A. Kiely}
\affiliation{Department of Physics, University College Cork, Cork, Ireland}

\author{A. Ruschhaupt}
\affiliation{Department of Physics, University College Cork, Cork, Ireland}

\begin{abstract}
Fast and robust quantum control protocols are often based on an idealised approximate description of the relevant quantum system. While this may provide a performance which is close to optimal, improvements can be made by incorporating elements of the full system representation. We propose a new technique for such scenarios, called enhanced shortcuts to adiabaticity (eSTA). The eSTA method works for previously intractable Hamiltonians by providing an analytical correction to existing STA protocols. This correction can be easily calculated and the resulting protocols are outside the class of STA schemes. We demonstrate the effectiveness of the method for three distinct cases: manipulation of an internal atomic state beyond the rotating wave approximation, transport of a neutral atom in an optical Gaussian trap and transport of two trapped ions in an anharmonic trap.
\end{abstract}
\maketitle

% ---------------------------------------------------------------
% Introduction
% ---------------------------------------------------------------

The development of quantum technologies for a wide variety of applications is a rapidly growing field \cite{schleichQuantumTechnologyResearch2016a}. However, a common roadblock is the requirement for fast and robust control of fragile quantum states, which is critical to exploiting any quantum advantage. The process must be fast to avoid long interaction times with the external environment (decoherence) and stable to avoid accumulation of errors. These problems have been addressed by a number of distinct techniques such as adiabatic methods \cite{vitanovStimulatedRamanAdiabatic2017a}, composite pulses \cite{torosovCompositePulsesUltrabroadband2015a,ivanovCompositeTwoqubitGates2015a,torosovHighfidelityErrorresilientComposite2014a}, numerical optimal control \cite{khanejaOptimalControlCoupled2005a,glaserTrainingSchrodingerCat2015a,vanfrankOptimalControlComplex2016a,doriaOptimalControlTechnique2011a} and shortcuts to adiabaticity 
\cite{torronteguiChapterShortcutsAdiabaticity2013a,guery-odelinShortcutsAdiabaticityConcepts2019b}.

Shortcuts to adiabaticity (STA) are analytical methods to design the time dependence of the Hamiltonian to ensure effective adiabatic state evolution in finite time. STA methods have the advantage of providing physical insight into the control process as well as constructing a whole class of protocols that achieve the desired result. In combination with perturbation theory, the optimal protocol in this class can be found which is most stable regarding a relevant type of noise or imperfection \cite{ruschhauptOptimallyRobustShortcuts2012a,ruschhauptShortcutsAdiabaticityTwolevel2014c,kielyInhibitingUnwantedTransitions2014b}. There has also been work to improve protocols in a non-perturbative manner \cite{impensFastQuantumControl2019a,levyNoiseResistantQuantum2018a}, using variational methods \cite{saberiAdiabaticTrackingQuantum2014,selsMinimizingIrreversibleLosses2017} and in combination with numerical optimal control \cite{corgierFastManipulationBose2018,mortensenFastStateTransfer2018a,campbellShortcutAdiabaticityLipkinMeshkovGlick2015, martinez-garaotFastQuasiadiabaticDynamics2015b}.

STA methods have been used to control a variety of Hamiltonians such as harmonic oscillator potentials \cite{chenFastOptimalFrictionless2010b,torronteguiFastAtomicTransport2011, kielyFastStableManipulation2015b} and two-level \cite{ruschhauptOptimallyRobustShortcuts2012a,kielyInhibitingUnwantedTransitions2014b}, three-level \cite{chenEngineeringFastPopulation2012,bensenySpatialNonadiabaticPassage2017b} and four-level \cite{kielyShakenNotStirred2016b} systems. They have been utilised experimentally for trapped ions \cite{anShortcutsAdiabaticityCounterdiabatic2016a}, superconducting qubits \cite{wangExperimentalRealizationHighfidelity2018, wangExperimentalRealizationFast2019}, nitrogen-vacancy centres \cite{zhangExperimentalImplementationAssisted2013, kolblInitializationSingleSpin2019}, ultracold atoms \cite{schaffShortcutsAdiabaticityTrapped2011} and Bose-Einstein condensates \cite{schaffShortcutAdiabaticityInteracting2011}. However there are still many Hamiltonians which are not tractable with standard STA techniques. Our new procedure is intended to deal with such cases.

In this paper we provide an analytical enhancement to STA protocols inspired by techniques of numerical optimal control, termed {\it enhanced shortcuts to adiabaticity} (eSTA). There are several key benefits. Firstly, the eSTA protocol provides higher fidelities than the STA protocol and may be outside the original class of STA protocols. This represents a significant improvement over previous methods based on optimisation inside the STA class \cite{ruschhauptOptimallyRobustShortcuts2012a}. Secondly, the resulting protocol is still completely analytical in nature, requiring no significant numerical computation or iterative procedure and therefore the resulting protocols can provide further physical insight. In addition, eSTA protocols can also serve as good initial seeds for further numerical optimisation (similar to recent attempts in utilising human intuition \cite{sorensenExploringQuantumSpeed2016}).
As STA methods have also been applied beyond quantum systems \cite{guery-odelinShortcutsAdiabaticityConcepts2019b}, for example in optical waveguides \cite{longhiQuantumopticalAnalogiesUsing2009,linModeConversionUsing2012,chungShortBroadbandSilicon2017}, classical mechanical systems 
\cite{dengBoostingWorkCharacteristics2013,deffnerClassicalQuantumShortcuts2014,gonzalez-resinesInvariantBasedInverseEngineering2017} and statistical physics 
\cite{guery-odelinNonequilibriumSolutionsBoltzmann2014}, eSTA has a broad range of applicability and in principle can be applied beyond the scope of quantum control.
In the following, we will outline the details of the eSTA method; the key result is summarised in Eq. \eqref{epsilon}. After this, we will demonstrate the flexibility of this approach by applying it to three different settings which are all ubiquitous in quantum technologies: population transfer in a two-level system beyond the rotating wave approximation \cite{scheuerPreciseQubitControl2014,londonStrongDrivingSingle2014,fuchsGigahertzDynamicsStrongly2009,ibanezPulseDesignRotatingwave2015}, transport of a single neutral atom in an optical trap \cite{ kuhrDeterministicDeliverySingle2001, beugnonTwodimensionalTransportTransfer2007, muldoonControlManipulationCold2012} and the transport of two trapped ions in an anharmonic trap \cite{waltherControllingFastTransport2012, bowlerCoherentDiabaticIon2012b, palmeroFastTransportTwo2013, palmeroFastTransportMixedspecies2014, luOptimalTransportTwo2015}.

\paragraph*{Formalism of Enhanced Shortcuts to Adiabaticity---} Consider a closed quantum system described by a Hamiltonian $\Hamil_{\mu}$, which we will refer to as the system Hamiltonian. Our goal is to change the Hamiltonian in time so that the system evolves from the initial state $\ket{ \Psi_0}$ at $t=0$ to the target state $\ket{ \Psi_T}$ in a given total time $t_f$. 
%
%
%%%%%%%%%%%%%%%%%%%%%%%%%%%%%%%%%%%%%%%%%%
%Figure 1
%%%%%%%%%%%%%%%%%%%%%%%%%%%%%%%%%%%%%%%%%%
\begin{figure}[t]
\begin{center}
\includegraphics[width=\linewidth]{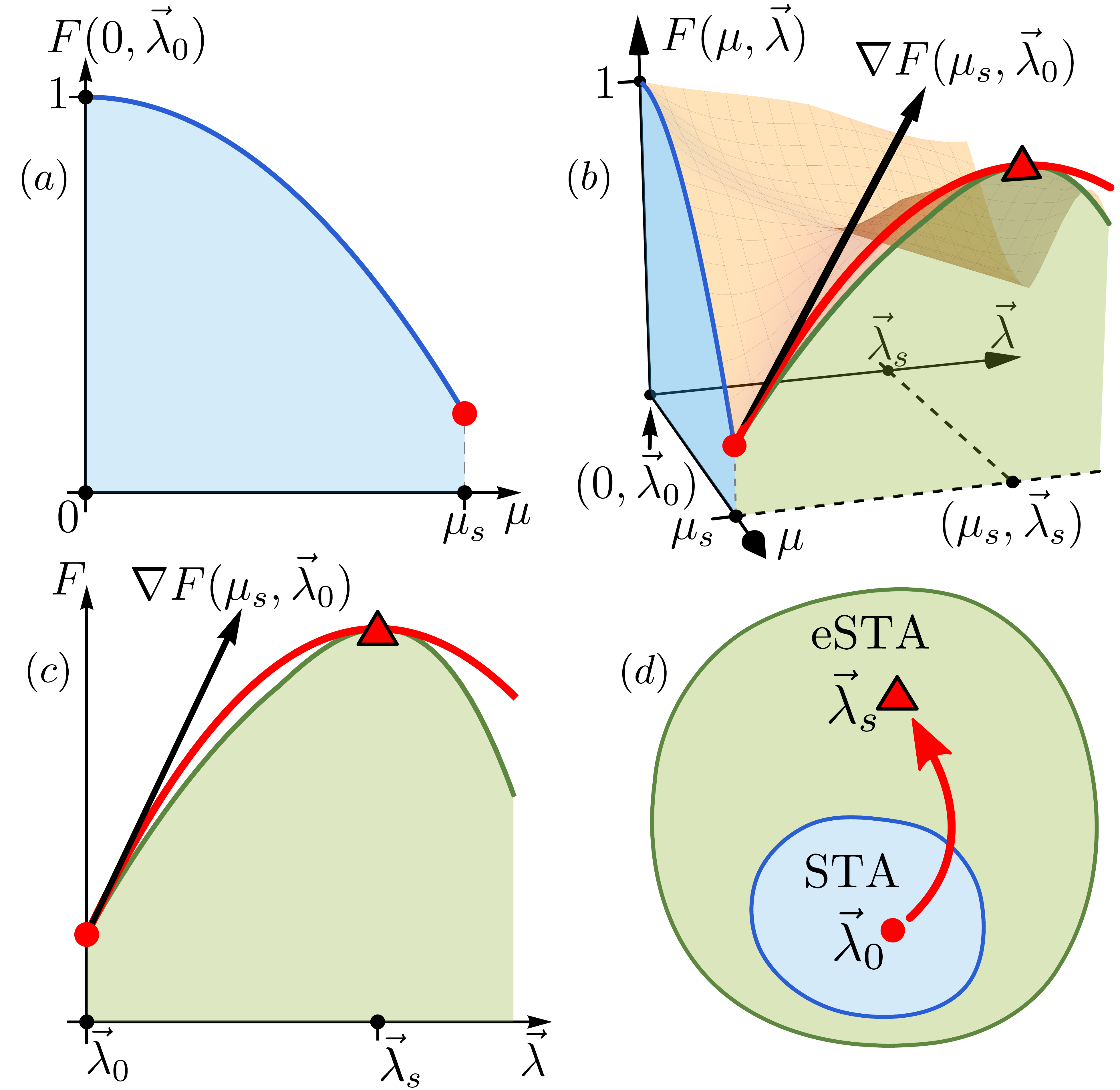}\;
\end{center}
\caption{\label{fig_1_schematic} (Colour online) Schematic overview of eSTA: (a) Fidelity using the STA protocol $\vec{\lambda}_0$ as a function of the Hamiltonian parameterisation $\mu$. Red dot indicates the fidelity at the point $\mu_s$. (b) Surface diagram of the fidelity for different Hamiltonians $\mu$ and different control protocols $\vec \lambda$. The black arrow shows the gradient at $(\mu_s,\vec{\lambda}_0)$. The solid red line indicates the parabolic approximation with the red triangle located at its maximum. (c) Cross section of part (b), showing the fidelity for the system Hamiltonian as a function of the control parameterisation. (d) Set of control protocols using STA methods (blue, inner region) and using eSTA (green, outer region)}
\end{figure}
%%%%%%%%%%%%%%%%%%%%%%%%%%%%%%%%%%%%%%%%%%
%
%
The Schr\"odinger equation is $i \hbar \frac{\partial}{\partial t} \ket{\Psi (t)} = \Hamil_\mu (\vec\lambda; t) \ket{\Psi (t)}$, where the value of $\mu$ fixes the form of the Hamiltonian and the time dependent control of the system parameters is characterised by $\vec\lambda$. The fidelity for this evolution is $\text{F}(\mu,\vec{\lambda})=
\fabsq{ \bra{\Psi_T}
	U_{\mu, \vec \lambda}(t_f,0) \ket{\Psi_0} }$.
First the system Hamiltonian $\Hamil_{\mu}$ (which is not easily dealt with using STA techniques) is approximated by an idealised, simpler Hamiltonian $\Hamil_{0}$ where an STA method can be applied. The manipulations required for the STA protocol are parameterised by $\vec{\lambda}_{0}\in \mathbb{R}^{N}$. Our goal is to find  $\vec{\lambda}$ such that the fidelity of this chosen evolution under $\Hamil_{\mu}$ is improved, where the method of improvement is motivated by the GRAPE (Gradient Ascent Pulse Engineering) algorithm \cite{khanejaOptimalControlCoupled2005a,wuOptimalSuppressionDefect2015}. 
Clearly, just using the STA protocol that was designed for the idealised Hamiltonian $\Hamil_0$ does not give perfect fidelity for the system Hamiltonian $\Hamil_{\mu}$ (see red dot in Fig. \ref{fig_1_schematic}(a)). However, we assume that the difference between the system and idealised Hamiltonians $\mu_s$ is small. Hence, we also assume that using the STA protocol $\vec{\lambda}_0$ for the system Hamiltonian is close to optimal, see Fig. \ref{fig_1_schematic}(b).
In order to calculate how much and in what manner to alter the original STA scheme $\vec{\lambda}_0$, we calculate the gradient with respect to $\vec{\lambda}$ and assume that the fidelity behaves quadratically in the neighbourhood of $(\mu_s,\vec{\lambda}_0)$, see Fig. \ref{fig_1_schematic}(c).

The new control function parameterised by $\vec\lambda$ is given by $\vec \lambda_{s} = \vec\lambda_0 + \vec\epsilon$, where the correction is
\begin{eqnarray}
\vec\epsilon \approx 
\frac{2\left[1-\text{F}(\mu_s,\vec \lambda_0)\right]}{\left|\nabla_{\vec \lambda} F(\mu_s,\vec \lambda_0)\right|}
\frac{\nabla_{\vec \lambda} F(\mu_s,\vec \lambda_0)}{\left|\nabla_{\vec \lambda} F(\mu_s,\vec \lambda_0)\right|}. \label{parab}
\end{eqnarray}
Here, we have assumed the fidelity at the maximum of the parabola (see red triangle in Figs. \ref{fig_1_schematic}(b) and \ref{fig_1_schematic}(c)) is approximately one i.e. $F(\mu_s,\vec{\lambda}_0+\vec{\epsilon})\approx 1$.

To calculate $\vec{\epsilon}$ we must estimate the gradient and the value of the fidelity at $(\mu_s,\vec{\lambda}_0)$, (see red dot in Fig. \ref{fig_1_schematic}).
To derive these estimates, we assume that the initial state $\ket{\Psi_0}$ and the final target state $\ket{\Psi_T}$ are independent
of the parameterisation $\mu$.  Since STA methods can be applied for the idealised Hamiltonian, the solutions $\ket{\chi_n (t)}$ are known (using Lewis-Riesenfeld invariants for example).
Since $U_{0,\vec \lambda_0}(t_2, t_1)$ is the time-evolution for $\mu=0$, we have that $\ket{\chi_n (t)} = U_{0,\vec \lambda_0} (t,0) \ket{\chi_n (0)}$ and $U_{0,\vec \lambda_0} (t,s) =\sum_{n} \ketbra{\chi_n(t)}{\chi_n(s)}$. We assume $\ket{\chi_0 (0)} = \ket{\Psi_0}$ and therefore $\ket{\chi_0 (t_f)} = \ket{\Psi_T}$.

We now estimate the terms needed to calculate the correction, assuming that we can neglect higher order contributions in both $\mu$ and $\vec\epsilon$. 
We start with a series expansion $\Hamil_\mu(\vec\lambda;t) = \sum_{n=0}^\infty \mu^n \Hamil^{(n)} (\vec\lambda;t)$ where $\Hamil^{(0)} (\vec\lambda;t)=\Hamil_{0} (\vec\lambda;t)$. Time-dependent perturbation theory \cite{QuantumMechanicsMessiah}, provides a series expansion of the corresponding time-evolution operator
$U_{\mu,\vec\lambda} (t_2,t_1) = \sum_{n=0}^\infty \mu^n U_{\vec\lambda}^{(n)} (t_2,t_1)$
where exact form of the first, second and third order can be found in \supplement.

From a series expansion of the fidelity in $\mu$ we get
$F(\mu,\vec\lambda_0) = \sum_{n=0}^\infty \mu^n F^{(n)}$.
The STA scheme works perfectly for the idealised Hamiltonian by construction, $F^{(0)} = 1$.
For the higher orders, one gets (see \supplement for details): $F^{(1)} = 0$,
$F^{(2)} = -\frac{1}{\hbar^2} \sum_{n=1}^\infty \fabsq{\int_0^{t_f} dt \, \alpha^{(1)}_{n,0} (t)}$,
and
$F^{(3)} \approx - \frac{2}{\hbar^2} \sum_{n=1}^\infty \mbox{Re} \left[\left(\int_0^{t_f} dt \, \alpha_{n,0}^{(1)}(t) \right)^*
\left(\int_0^{t_f} dt \, \alpha_{n,0}^{(2)}(t) \right)\right]$ where we have defined $\alpha^{(j)}_{n,m} (t) = \bra{\chi_n (t)} \Hamil^{(j)} (\vec\lambda_0; t) \ket{\chi_m (t)}$.
Using these results, the fidelity $F(\mu_s, \vec \lambda_0)$ can be approximated up to second order in $\mu_s$ as
$F(\mu_s, \vec \lambda_0)
\approx 1-\frac{1}{\hbar^2} \sum_{n=1}^\infty
\fabsq{\int_0^{t_f} dt \, \left(\mu_s \alpha^{(1)}_{n,0} (t) + \mu_s^2 \alpha_{n,0}^{(2)} (t) \right)}
\approx 1-\frac{1}{\hbar^2} \sum_{n=1}^\infty \fabsq{G_n}$,
where
\begin{eqnarray}
G_n = \!\! \int_0^{t_f} \!\! dt \braXket{\chi_n(t)}{\left[\Hamil_S (\vec\lambda_0;t) - \Hamil^{(0)} (\vec\lambda_0;t)\right]}{\chi_0(t)},
\label{eq_Gn}
\end{eqnarray}
and $\Hamil_S = \Hamil_{\mu_s}$.
The gradient of the fidelity with respect to $\vec\lambda$
can be expanded in $\mu$ as
$\nabla F (\mu, \lambda_0)
= \sum_{n=0}^\infty \mu^n \vec{\cF}^{(n)}$.
The relevant orders are in \supplement $\vec{\cF}^{(0)} = \vec{0}$,
%\begin{eqnarray}
$\vec{\cF}^{(1)} = 
-\frac{2}{\hbar^2} \sum_{n=1}^\infty \text{Re } \left[\int_0^{t_f} dt \, \alpha^{(1)}_{n,0} (t)
\int_0^{t_f} ds\, {\vec{\beta}^{(0)}_{n,0} (s)}^* \right],$
and
$\vec{\cF}^{(2)} \approx - \frac{2}{\hbar^2} \sum_{n=1}^\infty \mbox{ Re } \big[
\left(\int_0^{t_f} dt \, \vec{\beta}^{(1)}_{n,0}(t)\right)^* \left(\int_0^{t_f} dt \, \alpha^{(1)}_{n,0}(t)\right)\\
+ \left(\int_0^{t_f} dt \, \vec{\beta}^{(0)}_{n,0}(t)\right)^* \left(\int_0^{t_f} dt \, \alpha^{(2)}_{n,0}(t)\right)\big]$ where we have defined $\vec{\beta}^{(j)}_{n,m} (t)  = \bra{\chi_n (t)} \nabla \Hamil^{(j)} (\vec\lambda_0; t) \ket{\chi_m (t)}$.
From these results, we get up to second order in $\mu_s$ that
$\nabla F (\mu_s, \lambda_0) 
\approx
-\frac{2 \mu_s}{\hbar^2} \sum_{n=1}^\infty \text{Re} \Bigg\{ \int_0^{t_f} dt \left[\mu_s \alpha^{(1)}_{n,0} (t) + \mu_s^2 \alpha^{(2)}_{n,0} (t) \right]\\
\int_0^{t_f} ds\, \left[\vec{\beta}^{(0)}_{n,0} (s) + \mu \vec{\beta}^{(1)}_{n,0} (s)\right]^* \!\!\! \Bigg\}
\approx -\frac{2}{\hbar^2} \sum_{n=1}^\infty \text{Re} \left( G_n \vec{K}_n^* \right)$,
where
\begin{eqnarray}
\vec{K}_n &=& \int_0^{t_f} dt \, \bra{\chi_n (t)} \nabla \Hamil_S (\vec\lambda_0; t) \ket{\chi_0 (t)}.
\label{eq_Kn}
\end{eqnarray}

From Eq. \eqref{parab}, we finally arrive at the key result of the paper, the analytical expression for the eSTA protocol
\begin{eqnarray}
\vec\lambda_s \approx \vec\lambda_0-\frac{\left(\sum_{n=1}^N \fabsq{G_n}\right)\left[\sum_{n=1}^N \mbox{Re} \left(G_n^* \vec{K}_n\right)\right]}
{\fabsq{\sum_{n=1}^N \mbox{Re} \left(G_n^* \vec{K}_n\right)}},
\label{epsilon}
\end{eqnarray}
where $G_n$ is given by Eq. \eqref{eq_Gn}, $K_n$ is given by Eq. \eqref{eq_Kn}, and we have truncated the infinite sums to the first $N$ terms.

We underline that $G_n$ and $\vec{K}_n$ can be both easily calculated as only the Hamiltonians and the known solutions for the idealised Hamiltonian $\Hamil_0$ are involved. Note that using the eSTA method provides protocols which are outside the class of STA schemes (see Fig. \ref{fig_1_schematic}(d)) and so represents a significant improvement over previous perturbation based optimisation \cite{ruschhauptOptimallyRobustShortcuts2012a}.

% -----------------------------------------------------------------------------
% Population inversion without the Rotating-Wave-Approximation
% -----------------------------------------------------------------------------

\paragraph*{Population inversion without the Rotating-Wave-Approximation---}
As a first example, we consider the following system Hamiltonian
\begin{eqnarray}
\Hamil_S = \frac{\hbar}{2} \left(\begin{array}{cc} -\delta(\vec{\lambda};t) & \Omega^* (\vec{\lambda};t) \left(1 + e^{-2i\omega t} \right) \\ \Omega (\vec{\lambda};t) \left(1 + e^{2i\omega t} \right) & \delta(\vec{\lambda};t) \end{array}\right), \label{Ham2lvl} \nonumber \\
\end{eqnarray}
which generically appears in many areas of quantum technologies. A common setting is that of two atomic states which are coupled by a classical light source (e.g. a laser) where the Rabi frequency $\Omega$ depends on the light amplitude and the detuning $\delta$ depends on the light frequency $\omega$.

The terms $e^{\pm 2i\omega t}$ are typically neglected, which is known as the ``Rotating-wave-approximation"(RWA) \cite{allenOpticalResonanceTwolevel1987}.
Our idealised Hamiltonian $\Hamil_0$ is then just $\Hamil_S$ where the terms $e^{\pm 2i\omega t}$ are set to zero.

While this approximation may work well for adiabatic methods, it will fail for fast nonadiabatic operations. Our goal is to use the eSTA method to provide fast population inversion even in the regime where the RWA does not hold (i.e. small values of $t_f$). This has been previous attempted using numerical methods \cite{ibanezPulseDesignRotatingwave2015}; however here it will be done analytically.

Our initial scheme $\vec{\lambda}_0$ was derived to be stable concerning systematic errors in the Rabi frequency \cite{ruschhauptOptimallyRobustShortcuts2012a} (e.g. arising from the Gaussian profile of the laser), and is given by
$\Omega (\vec{\lambda}_0;t) = \frac{\pi}{t_f} \sqrt{1+16 \sin\left(\frac{\pi t}{t_f}\right)^6}$,
$\delta (\vec{\lambda}_0;t) = -8 \frac{\pi}{t_f}\sin\left(\frac{\pi t}{t_f}\right)\sin\left(\frac{2 \pi t}{t_f}\right) [1+4 \sin\left(\frac{\pi t}{t_f}\right)^6]/[1+16 \sin\left(\frac{\pi t}{t_f}\right)^6]$.
By design, this scheme gives perfect population inversion for the idealised Hamiltonian $\Hamil_{0}$.

The scheme is modified as $\Omega (\vec{\lambda};t)= \Omega (\vec{\lambda}_0;t)+ f_1(\vec{\epsilon};t)$ and $\delta (\vec{\lambda};t) = \delta (\vec{\lambda}_0;t)+ f_2(\vec{\epsilon};t)$ where $f_1$ and $f_2$ are the minimal polynomial functions which fulfil $f_i(\vec{\epsilon};t')=0$ for $t'=0, t_f$ and $f_i(\vec{\epsilon};\frac{j t_f}{5})=\epsilon_{4(i-1)+j}$, where we have chosen to use $8$ components $\vec{\epsilon}=(\epsilon_n)_{n=1,\ldots,8}$. Note that $f_1(\vec{0},t)=0$ and $f_2(\vec{0},t)=0$.
Since there are only two solutions in this setting, we can calculate $\vec{\epsilon}$ exactly, without any truncation. These solutions can be found analytically using Lewis-Riesenfeld Invariants \cite{ruschhauptOptimallyRobustShortcuts2012a}.

%%%%%%%%%%%%%%%%%%%%%%%%%%%%%%%%%%%%%%%%%%
%Figure 2
%%%%%%%%%%%%%%%%%%%%%%%%%%%%%%%%%%%%%%%%%%
\begin{figure}[t]
	\begin{center}
		\includegraphics[width=0.98\linewidth]{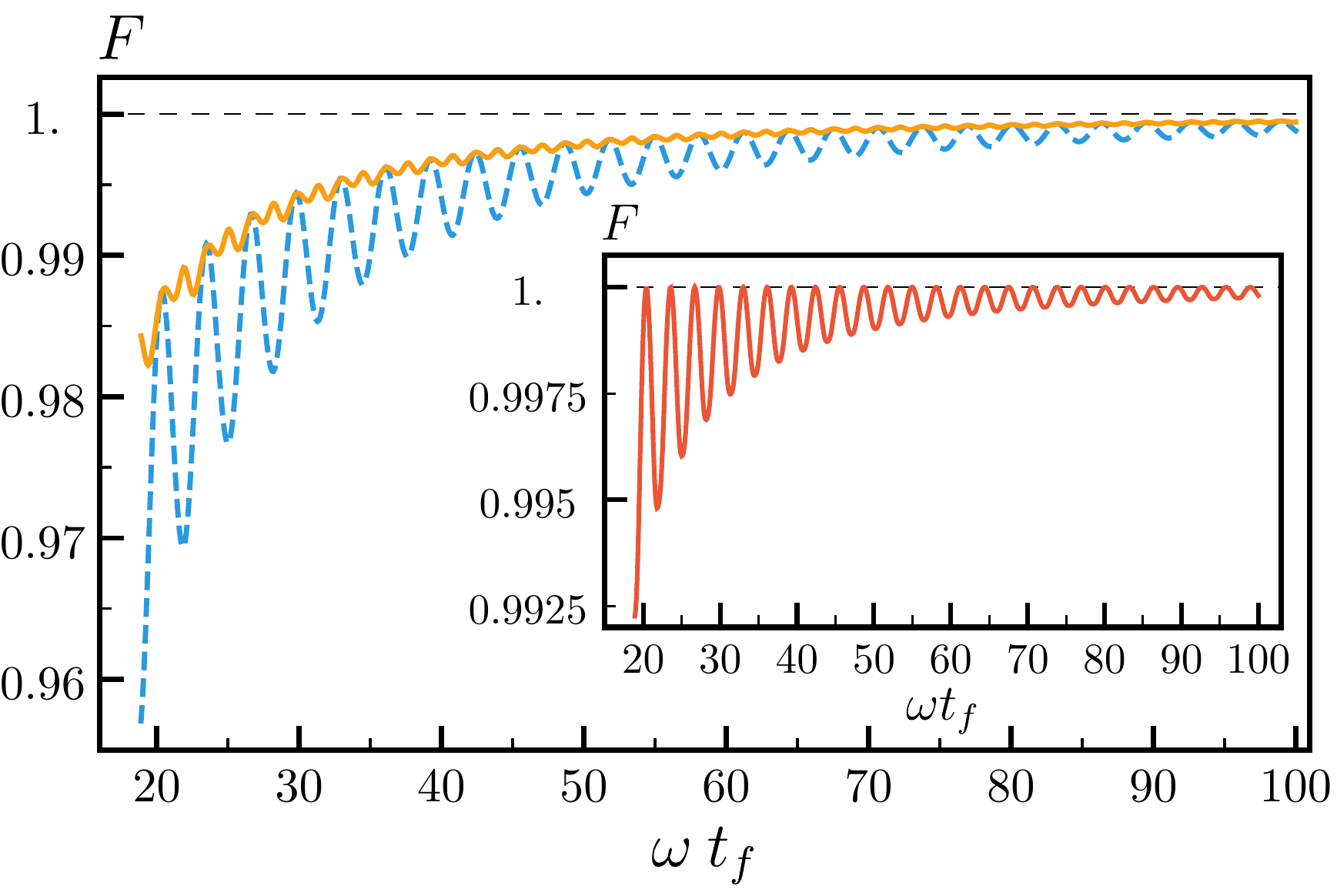}
	\end{center}
	\caption{\label{fig_2_RW} (Colour online) Population inversion without rotating-wave approximation: Fidelity $F$ versus final time $t_f$ for two-level Hamiltonian Eq. \eqref{Ham2lvl}; STA scheme $\vec{\lambda}_0$ (blue, dashed line) and the eSTA scheme $\vec{\lambda}_s$ (orange, solid line). Inset: Fidelity $F$ for the idealised Hamiltonian $\Hamil_0$ using the eSTA scheme $\vec{\lambda}_s$ (red, solid line).}
\end{figure}
%%%%%%%%%%%%%%%%%%%%%%%%%%%%%%%%%%%%%%%%%%

The fidelity using the STA scheme and eSTA schemes for the system Hamiltonian $\Hamil_S$, is shown in Fig. \ref{fig_2_RW}. For the shown final times $t_f$, the eSTA scheme outperforms the original STA scheme, since it always results in a higher or equal fidelity.
The eSTA schemes are outside the set of STA control functions (see Fig. \ref{fig_1_schematic}(d)). This can been seen by calculating the fidelity of the eSTA schemes for the the idealised Hamiltonian  $\Hamil_0$ (see inset of Fig.\ref{fig_2_RW}). Since applying the STA scheme to the idealised Hamiltonian gives unit fidelity for all total times $t_f$ by construction, every fidelity value below one shows that the eSTA scheme is outside the set of STA schemes.

% ------------------------------------------------------------------------
% ------------------------ One particle transport ------------------------
% ------------------------------------------------------------------------

\paragraph*{Single Particle Transport---}
We consider transport of a particle in a one dimensional trap over a distance $d$ in a total time $t_f$. The trap trajectory $q_0(\vec{\lambda},t)$ is parameterised by a real valued control vector $\vec{\lambda}=(\lambda_1,...,\lambda_6)$, so that $q_0(0)=0$, $q_0 (t_f)=d$ and $q_0(j/7)=\lambda_j$ for $j=1,...,6$.
The system/idealised Hamiltonian is $\Hamil_{S/0}=\frac{p^2}{2 m} + V_{S/0} [x-q_0(\vec{\lambda},t)]$ where $V_S (x)
= U_0
\left[
1-\exp\left( -\frac{m \omega^2}{2 U_0} x^2\right)
\right]$ is a Gaussian potential, and $V_{0} (x)=\frac{1}{2} m \omega^2 x^2,$ since $V (x) \to V_0 (x)$  for $\mu=1/a \to 0$ where $a = U_0/(\hbar \omega)$.
There are known STA techniques for $\Hamil_0$ to design trajectories $q_0$ that give perfect fidelity e.g. using Lewis-Riesenfeld invariants \cite{torronteguiFastAtomicTransport2011}.
A known dynamical invariant for harmonic trap transport has the form $I(t)=\frac{1}{2m}\left( p - m \dot{q}_c \right)^2
+\frac{1}{2}m\omega_0^2 \left[ x - m q_c\left(t\right) \right]^2$ where $q_c(t)$ must satisfy the auxiliary equation
\begin{align}\label{equ:aux}
\ddot{q}_c+\omega_0^2\left(q_c - q_0 \right) = 0.
\end{align}
This equation relates the physical trap trajectory $q_0(t)$ with the particle's classical path $q_c(t)$ (which parameterises the state). $q_0(t)$ can be inverse engineered using boundary conditions and an appropriately chosen $q_c(t)$ via Eq. \eqref{equ:aux}. To ensure the system is in the groundstate after transport and that the trap is stationary, we require the boundary conditions
$q_c(0) = 0$, $q_c(t_f)= 0$, $\frac{d^n q_c(t')}{dt^n}=0$ for $n=1,\ldots,4$ at $t'=0,t_f$.
We set $q_c(t)=\sum_{n=0}^{10} c_n t^n$, and find $q_0(t)$ from Eq. \eqref{equ:aux}.
To implement eSTA we need only to calculate $\vec{\epsilon}$ using $G_n$ and $\vec{K}_n$. The eigenstates $\ket{\chi_n (s)}$ are known analytically from \cite{torronteguiFastAtomicTransport2011}, and so the  integrals $G_n$ and $K_n$ can be calculated for each $n$. In the following, we will show that using $N=1$ is sufficient.

%%%%%%%%%%%%%%%%%%%%%%%%%%%%%%%%%%%%%%%%%%
%Figure 3
%%%%%%%%%%%%%%%%%%%%%%%%%%%%%%%%%%%%%%%%%%
\begin{figure}[t]
\includegraphics[width=0.98\linewidth]{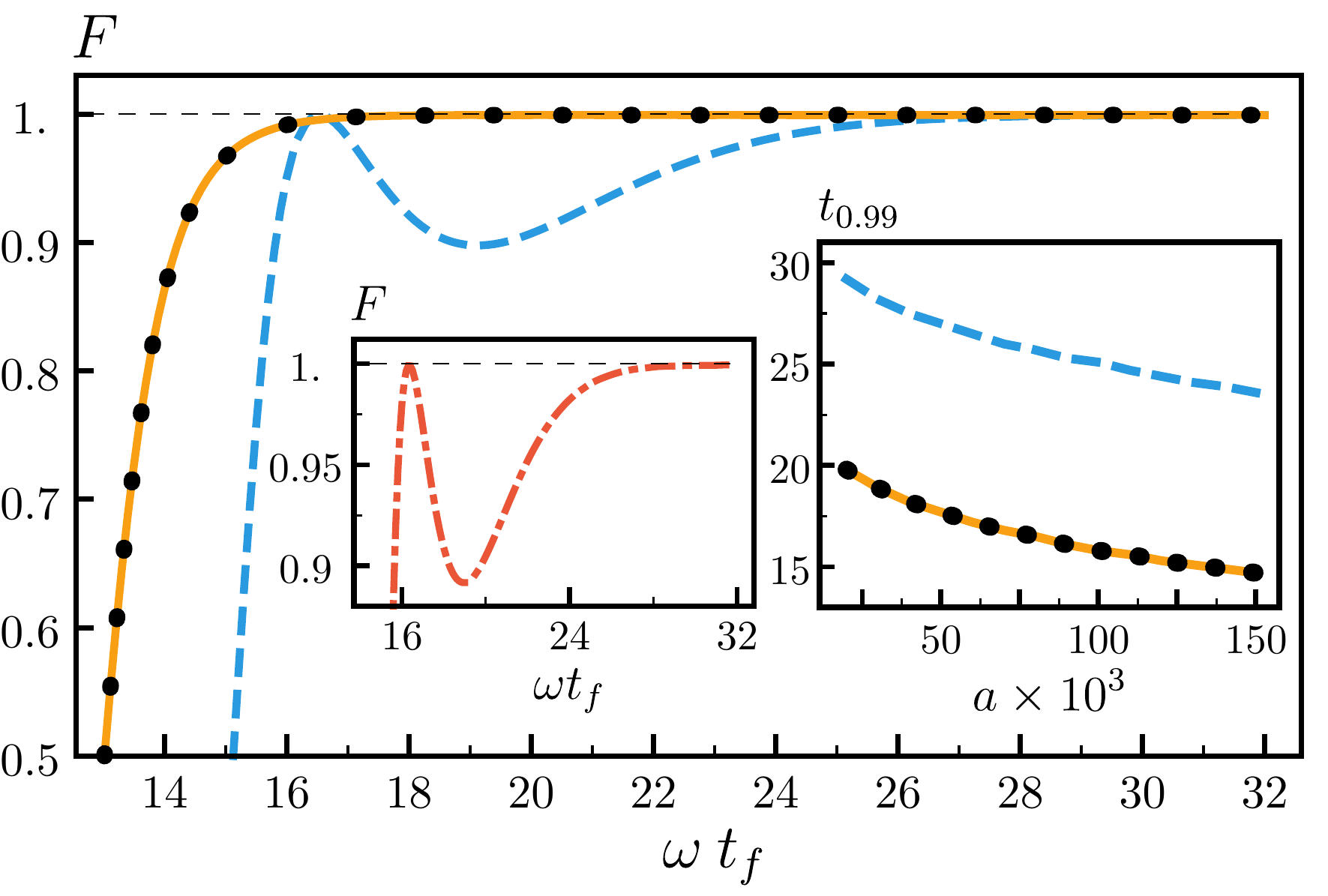}
\caption{(Colour online) Transport of a single particle: Fidelity $F$ versus total time $t_f$ using the STA scheme $\vec{\lambda}_0$ (blue, dashed line) and the eSTA scheme $\vec{\lambda}_s$ ($N=1$: orange, solid line, $N=2$: black, dots); $a = 100 \times 10^3$. Left inset: Fidelity $F$ for the idealised Hamiltonian $\Hamil_0$ using the eSTA scheme $\vec{\lambda}_s$ (red, dashed-dotted line). Right inset: threshold time $t_{0.99}$ versus $a$. Transport distance $d=1562 \sigma$.
\label{fig_3_transport_one_particle}}
\end{figure}
%%%%%%%%%%%%%%%%%%%%%%%%%%%%%%%%%%%%%%%%%%

In Fig. \ref{fig_3_transport_one_particle}, the fidelity $F$ is shown versus different final times $t_f$ using the STA transport scheme $\vec{\lambda}_0$ (blue, dashed line) and eSTA transport scheme $\vec{\lambda}_s$ (using $N=1$, orange, solid line). This was calculated numerically where the time evolution was performed using the Fourier split-operator method and the initial ground state was found by imaginary-time evolution. For generality, we use  natural units; the frequency $\omega$ of the approximated harmonic potential as the inverse time unit, $\sigma=\sqrt{\hbar/(m\omega)}$ as the length scale and $\hbar\omega$ as the energy scale. We set the transport distance to $d=1562 \sigma$ and $a = 100 \times 10^3$.
The chosen dimensionless values can correspond to different physical settings, for example to a
$^{87}$Rb atom within an optical Gaussian trap of $U_0 = 0.4\,\mbox{mK}$, a trap width of $w = 2 \sqrt{a} \sigma = 334 \,\mu$m (purposely chosen as very wide to be far from the regime of classical and adiabatic motion) and a transport distance of $d = 825 \, \mu$m.

We see a significant improvement in transport fidelity using eSTA in comparison with STA. For extremely short times the approximation breaks down and neither STA nor eSTA produce good fidelity. For longer times the system approaches adiabaticity and the two schemes converge.
Clearly it is sufficient to consider just the first order, as the results for $N=1$ and $N=2$ are identical, see Fig. \ref{fig_3_transport_one_particle}. While not a requirement, we note that the symmetry of the STA trajectory is preserved by the eSTA protocol.

To highlight these eSTA schemes are outside the set of STA schemes (Fig. \ref{fig_1_schematic}(d)), we calculate the fidelity using $\lambda_s$ for the idealised Hamiltonian $\Hamil_0$ (left inset in Fig. \ref{fig_3_transport_one_particle}, red dashed-dotted line). By design every STA scheme must give a fidelity of exactly one, which is not necessarily true when using the eSTA protocol.

To examine the dependence on $a$, we look at a threshold time $t_{0.99}$ which is defined as the time such that the fidelity $F \ge 0.99$ for all final times $t_f \ge t_{0.99}$. The right inset plot shows this threshold time $t_{0.99}$ versus different values of $a$. We see that the eSTA threshold time (orange, solid line) decreases with increasing $a$ and is always much lower than the corresponding STA threshold time (blue, dashed line).
We have also investigated other potentials which produced qualitatively similar results to the Gaussian trap. This underlines the broad applicability of eSTA for single particle transport.

% ---------------------------------------------------------------
% ---- Transport of two ions including Coulomb interaction ------
% ---------------------------------------------------------------

\paragraph*{Transport of two ions including Coulomb interaction---}
We now consider transport of two interacting (Coulomb) ions with equal mass $m$ and charge $+e$, in a one dimensional Gaussian trap $V_S$ (as in the previous case), over a distance $d$, and in total time $t_f$.
The coordinates of the ions in the lab frame are given by $x_1$ and $x_2$. We assume that $x_1>x_2$ and  treat the ions as distinguishable.
We define $M=2m$ and move to centre-of-mass and relative coordinates defined by $x_{\text{c}}= \left( x_1 + x_2 \right)/2$ and
$x_{\text{r}}= \left( x_1 - x_2 \right)/2$.
The system/idealised Hamiltonian then becomes
$
\Hamil_{S/0}
=
\frac{p_\text{c}^2}{2 M}+ \frac{p_\text{r}^2}{2 M} + \frac{C}{2 x_\text{r}} + V_{S/0} \left[x_\text{c} + x_\text{r}-q_0(\vec{\lambda},t)\right]
+ V_{S/0} \left[x_\text{c} - x_\text{r}-q_0(\vec{\lambda},t)\right]$
where $C=e^2 / 4 \pi \epsilon_0$ and  
$V_0(x)=\frac{1}{2} M \omega^2 x^2$ in this case.
As $\Hamil_{0}$ becomes separable in centre-of-mass and relative coordinates, STA techniques need only be applied to the centre-of-mass part \cite{palmeroFastTransportTwo2013}.
%%%%%%%%%%%%%%%%%%%%%%%%%%%%%%%%%%%%%%%%%%
%Figure 4
%%%%%%%%%%%%%%%%%%%%%%%%%%%%%%%%%%%%%%%%%%
\begin{figure}[t]
\includegraphics[width=0.98\linewidth]{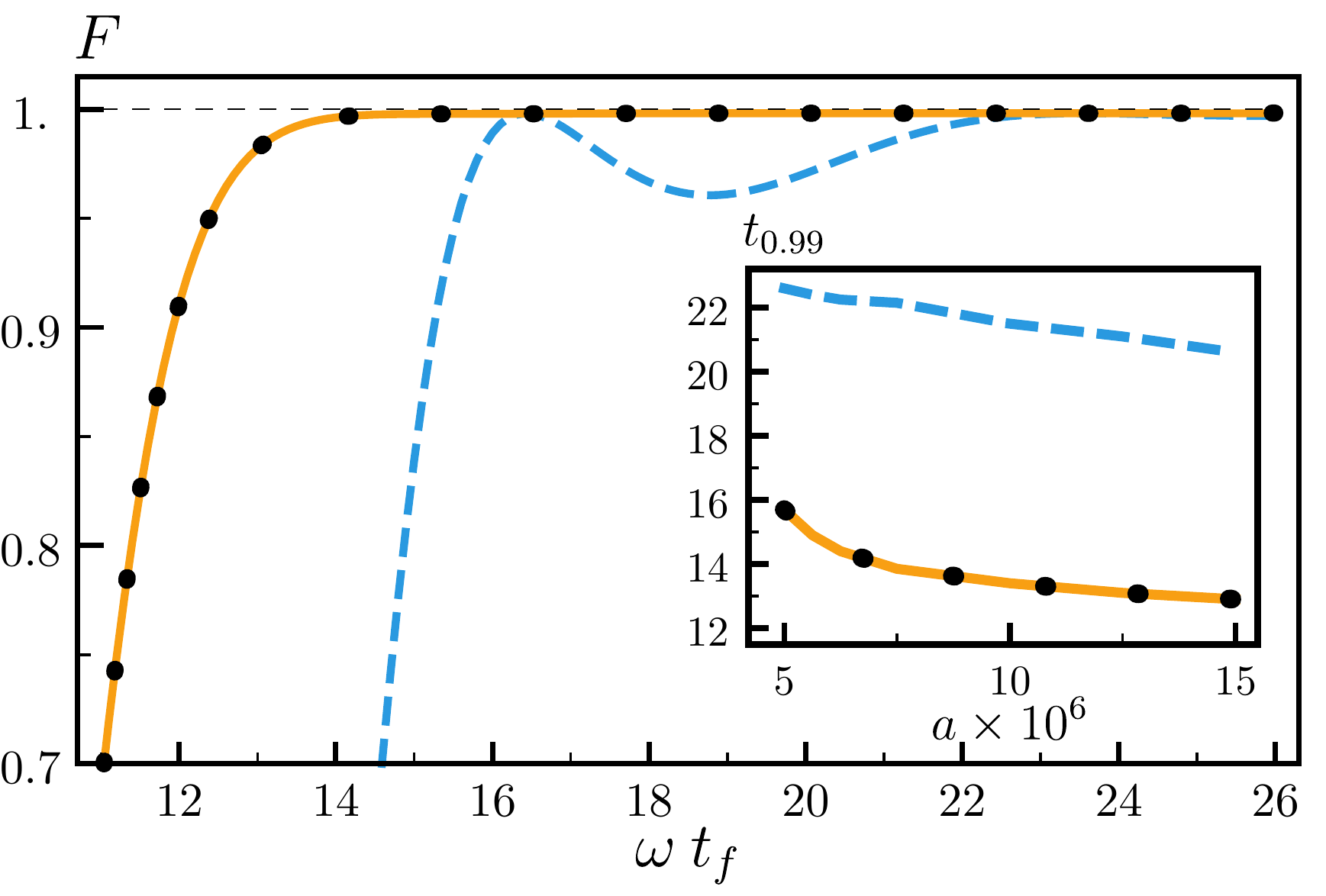}
\caption{(Colour online) Transport of two ions including Coulomb interaction: $a=10^7$ for the outer figure, $\tilde C= 7.35 \times 10^7$;
otherwise see caption of Fig. \ref{fig_3_transport_one_particle}.
\label{fig_transport_2_particles}}
\end{figure}
%%%%%%%%%%%%%%%%%%%%%%%%%%%%%%%%%%%%%%%%%%
For the eSTA scheme, we assume that the relative distance between the ions is constant and equal to the stationary equilibrium distance when calculating $G_n$ and $\vec{K}_n$.

In Fig. \ref{fig_transport_2_particles}, the fidelity $F$ is shown versus different final times $t_f$ using the STA transport scheme $\vec{\lambda}_0$ (dashed, blue line) and the eSTA transport scheme $\vec{\lambda}_s$ (using $N=1$, orange, solid line). We set the transport distance to $d=1562 \sigma$, $a = 10^7$ and the dimensionless Coulomb constant $\tilde C = \frac{e^2}{4\pi\epsilon_0} \frac{1}{\sigma\hbar\omega} = 7.35\times 10^7$. These dimensionless values can correspond to different physical settings, for example transport of two $^{9}$Be$^{+}$ ions over 370 $\mu$m in a surface ion trap of depth $U_0 = 0.8$ meV,
and frequency $0.13$ MHz (where the trap width has been chosen again to be large).

Similar to the previous case,
we see a significant improvement in transport fidelity based on eSTA in comparison with STA. For longer times the system approaches adiabaticity and the two schemes converge. It is sufficient to consider just the first order (i.e. set $N=1$ when calculating the eSTA scheme); to check this we also plot the result for $N=2$ (black dots) which give identical results. As before, one finds that the eSTA schemes are outside the set of STA schemes. The inset of Fig. \ref{fig_transport_2_particles} shows the  threshold time $t_{0.99}$ versus $a$.
We see that the eSTA threshold times (using $N=1$, orange, solid line) decreases with increasing $a$ and it is always significantly lower than the corresponding STA threshold time (blue dashed line).

% ------------------------------------------------------------------------
% -------------- Conclusion ----------------------------------------------
% ------------------------------------------------------------------------

\paragraph*{Conclusion---}In this paper we have presented an analytic extension to previous STA quantum control methods. We have demonstrated through three complementary examples relevant to quantum technologies, that this method can be applied to improve performance and also can be used to achieve physical insight.
Further work could focus on deriving strict criteria and uncertainty relations for when the method works effectively.
The eSTA procedure could be extended in several ways such as to condensates, open systems, and even beyond the scope of quantum control. 

We are grateful to D. Rea for useful discussion and
commenting on the manuscript. C.W. acknowledges support
by the Irish Research Council (GOIPG/2017/1846).

%----------------------------------------
\begin{appendix}

\begin{widetext}
\section{Estimations of $F(\mu,\vec\lambda_0)$ and $\nabla F (\mu, \vec{\lambda}_0)$ \label{app}}
In the following, we will provide further details concerning the estimations of
$F(\mu,\vec\lambda_0)$ and $\nabla F (\mu, \vec{\lambda}_0)$ which are used to derive the main
formula of enhanced Shortcuts to Adiabaticity (eSTA). 
A series expansion of the time-evolution operator of the system Hamiltonian is
\begin{eqnarray}
U_{\mu,\vec\lambda} (t_2,t_1) = \sum_{n=0}^\infty \mu^n U_{\vec\lambda}^{(n)} (t_2,t_1),
\label{seriesU}
\end{eqnarray}
where the first order is
\begin{eqnarray}
U_{\vec\lambda}^{(1)} (t_2,t_1) = -\frac{i}{\hbar} \int_{t_1}^{t_2} dt\, U_{\vec\lambda}^{(0)} (t_2, t)
\Hamil^{(1)} (\vec\lambda;t)
U_{\vec\lambda}^{(0)} (t, t_1),%\nonumber\\
\label{U_first}
\end{eqnarray}
the second order is
\begin{eqnarray}
U_{\vec\lambda}^{(2)} (t_2,t_1) &=& -\frac{1}{\hbar^2} \int_{t_1}^{t_2} dt \int_{t_1}^{t} ds \,
U_{\vec\lambda}^{(0)} (t_2, t)
\Hamil^{(1)} (\vec\lambda;t)
U_{\vec\lambda}^{(0)} (t, s)
\Hamil^{(1)} (\vec\lambda;s)
U_{\vec\lambda}^{(0)} (s, t_1)\nonumber\\
& & - \frac{i}{\hbar} \int_{t_1}^{t_2} dt\,
U_{\vec\lambda}^{(0)} (t_2, t)
\Hamil^{(2)} (\vec\lambda;t)
U_{\vec\lambda}^{(0)} (t, t_1),
\label{U_second}
\end{eqnarray}
and finally the third order is
\begin{eqnarray}
U_{\vec\lambda}^{(3)} (t_2,t_1) &=&
\frac{i}{\hbar^3} \int_{t_1}^{t_2} dt \int_{t_1}^{t} ds \int_{t_1}^{s} du %\nonumber\\
\ U_{\vec\lambda}^{(0)} (t_2, t)
\Hamil^{(1)} (\vec\lambda;t)
U_{\vec\lambda}^{(0)} (t, s)
\Hamil^{(1)} (\vec\lambda;s)
U_{\vec\lambda}^{(0)} (s, u)
\Hamil^{(1)} (\vec\lambda;u)
U_{\vec\lambda}^{(0)} (u, t_1) \nonumber\\
&-&\frac{1}{\hbar^2} \int_{t_1}^{t_2} dt \int_{t_1}^{t} ds %\nonumber\\
\ U_{\vec\lambda}^{(0)} (t_2, t)
\left[\Hamil^{(1)} (\vec\lambda;t)U_{\vec\lambda}^{(0)} (t, s)\Hamil^{(2)} (\vec\lambda;s)
+ \Hamil^{(2)} (\vec\lambda;t)U_{\vec\lambda}^{(0)} (t, s)\Hamil^{(1)} (\vec\lambda;s) \right]
U_{\vec\lambda}^{(0)} (s, t_1) \nonumber\\
&-& \frac{i}{\hbar} \int_{t_1}^{t_2} dt\,
U_{\vec\lambda}^{(0)} (t_2, t)
\Hamil^{(3)} (\vec\lambda;t)
U_{\vec\lambda}^{(0)} (t, t_1).
\label{U_third}
\end{eqnarray}
\end{widetext}
We define the following useful matrix elements
\begin{eqnarray}
\alpha^{(j)}_{n,m} (t) &=& \bra{\chi_n (t)} \Hamil^{(j)} (\vec\lambda_0; t) \ket{\chi_m (t)}, \label{alpha}\\
\vec{\beta}^{(j)}_{n,m} (t)  &=& \bra{\chi_n (t)} \nabla \Hamil^{(j)} (\vec\lambda_0; t) \ket{\chi_m (t)}.  \label{beta}
\end{eqnarray}
Note also that the matrix elements defined in Eq. \eqref{alpha} and Eq. \eqref{beta} obey the relations $\alpha^{(j)}_{n,m} (t) = {\alpha^{(j)}_{m,n} (t)}^*$ and
$\vec{\beta}^{(j)}_{n,m} (t) = {\vec{\beta}^{(j)}_{m,n} (t)}^*$. Hence it follows that $\alpha^{(j)}_{n,n} (t)$ and $\vec{\beta}^{(j)}_{n,n} (t)$ are real. 

\subsection{Approximation of $F(\mu,\vec\lambda_0)$}
From a series expansion of the fidelity in $\mu$ we get
\begin{eqnarray}
F(\mu,\vec\lambda_0) = \sum_{n=0}^\infty \mu^n F^{(n)}.
\label{app_fid}
\end{eqnarray}
By defining
\begin{eqnarray}
u_j = \braXket{\chi_0 (t_f)}{U_{\vec\lambda_0}^{(j)} (t_f,0)}{\chi_0 (0)},
\end{eqnarray}
we can express the coefficients in Eq. \eqref{app_fid} generally as
\begin{eqnarray}
F^{(n)}&=& \sum_{k=0}^n u_{n-k} u_k^* \nonumber \\
		&=& \begin{cases}
		 \fabs{u_{n/2}}^2+2 \mbox{ Re }  \left(\sum_{k=0}^{\frac{n}{2}-1} u_{n-k} u_k^*\right), & \text{ $n$ is even} \\
		2 \mbox{ Re }  \left(\sum_{k=0}^{(n-1)/2} u_{n-k} u_k^*\right), & \text{ $n$ is odd}
		\end{cases} \nonumber \\
\end{eqnarray}

% --- u_0 and F0 ------
Since the STA scheme works perfectly for the idealised Hamiltonian by construction, we have that $u_0 = 1$ and hence $F^{(0)} = 1$.

% --- u_1 and F1 -----
From Eq. \eqref{U_first}, we get 
\begin{eqnarray}
u_1
&=& -\frac{i}{\hbar} \int_{0}^{t_f} dt \, \bra{\chi_0 (t)} \Hamil^{(1)} (\vec\lambda_0; t)\ket{\chi_0 (t)}\nonumber\\
&=& -\frac{i}{\hbar} \int_{0}^{t_f} dt \, \alpha^{(1)}_{0,0} (t).
\end{eqnarray}
As $u_1$ is purely imaginary, it follows that $F^{(1)} = 0$.

% --- u2 and F2 ----
Now by using Eq. \eqref{U_second}, we get
\begin{eqnarray}
u_2 &=& - \frac{1}{\hbar^2} \int_0^{t_f} dt \int_0^{t} ds \nonumber\\
& & \quad \bra{\chi_0 (t)}
\Hamil^{(1)} (\vec\lambda;t)
U_{\vec\lambda_0}^{(0)} (t, s)
\Hamil^{(1)} (\vec\lambda;s)
\ket{\chi_0 (s)}\nonumber\\
& & -\frac{i}{\hbar} \int_{0}^{t_f} dt \, \bra{\chi_n (t)} \Hamil^{(2)} (\vec\lambda_0; t)\ket{\chi_0 (t)}.
\end{eqnarray}
Using $U_{\vec\lambda_0}^{(0)} (t, s)=\sum_n \ketbra{\chi_n (t)}{\chi_n (s)}$
simplifies this to
\begin{eqnarray}
u_2 &=& - \frac{1}{\hbar^2} \int_0^{t_f} dt \int_0^{t} ds
\sum_n \alpha^{(1)}_{0,n} (t) \alpha^{(1)}_{n,0} (s) \nonumber\\
& & -\frac{i}{\hbar} \int_{0}^{t_f} dt \, \alpha^{(2)}_{0,0} (t).
\end{eqnarray}
After a suitable transformation of the integration variables, we can also write this as
\begin{eqnarray}
u_2 &=& - \frac{1}{2 \hbar^2} \int_0^{t_f} dt \sum_n \nonumber\\
& & \quad \left[\int_0^{t} ds\,
\alpha^{(1)}_{0,n} (t) \alpha^{(1)}_{n,0} (s)  
+ \int_t^{t_f} ds\, \alpha^{(1)}_{0,n} (s) \alpha^{(1)}_{n,0} (t)\right]
\nonumber\\
& & -\frac{i}{\hbar} \int_{0}^{t_f} dt \, \alpha^{(2)}_{0,0} (t).
\end{eqnarray}
This form will be useful for calculating $F^{(2)}$ since
\begin{eqnarray}
2 \text{ Re } (u_2) &=&  - \frac{1}{\hbar^2} \sum_n \int_0^{t_f} dt \int_0^{t_f} ds \, \text{Re} \left[\alpha^{(1)}_{0,n} (t) \alpha^{(1)}_{n,0} (s)\right] \nonumber\\
&=& - \frac{1}{\hbar^2} \sum_n \text{Re} \left[ 
\int_0^{t_f} dt \, \alpha^{(1)}_{0,n} (t)
\int_0^{t_f} ds \, \alpha^{(1)}_{n,0} (s)\right]\nonumber\\
&=& - \frac{1}{\hbar^2} \sum_n \fabsq{\int_0^{t_f} dt \, \alpha^{(1)}_{n,0} (t)}
\end{eqnarray}
because $\alpha^{(1)}_{0,n}={\alpha^{(1)}_{n,0}}^* $ and $\alpha^{(2)}_{0,0} (t)$ is real.
Finally, we get
\begin{eqnarray}
F^{(2)} &=& -\frac{1}{\hbar^2} \sum_{n=1}^\infty \fabsq{\int_0^{t_f} dt \, \alpha^{(1)}_{n,0} (t)}.
\end{eqnarray}

% --- u3 and F3 ----
In a similar way by using Eq. \eqref{U_third}, we arrive at
\begin{eqnarray}
\lefteqn{u_3 = 
\frac{i}{\hbar^3} \int_{t_1}^{t_f} dt \int_{0}^{t} ds \int_{0}^{s} du
\sum_{n,m} \alpha_{0,n}^{(1)} (t) \alpha_{n,m}^{(1)} (s) \alpha_{n,0}^{(1)}(u)} & & \nonumber\\
\quad & & -\frac{1}{\hbar^2} \int_{0}^{t_f} dt \int_{0}^{t} ds
 \sum_n \left[\alpha_{0,n}^{(1)} (t) \alpha_{n,0}^{(2)} (s)
+  \alpha_{0,n}^{(2)} (t) \alpha_{n,0}^{(1)} (s) \right]\nonumber\\
& & - \frac{i}{\hbar} \int_{0}^{t_f} dt\, \alpha_{0,0}^{(3)}.
\end{eqnarray}
We use this together with previous results to calculate $F^{(3)}$. However,
we restrict to the contributions which involve double integrals
(while ignoring the contributions with three integrals over time). In such a way,
\begin{eqnarray*}
F^{(3)} \approx - \frac{2}{\hbar^2} \sum_{n=1}^\infty \mbox{Re} \left[\left(\int_0^{t_f} dt \, \alpha_{n,0}^{(1)}(t) \right)^*
\left(\int_0^{t_f} dt \, \alpha_{n,0}^{(2)}(t) \right)\right].
\end{eqnarray*}

\subsection{Approximation of $\nabla F (\mu, \vec{\lambda}_0)$}
The gradient of the fidelity 
\begin{eqnarray}
\nabla F (\mu, \vec{\lambda}_0) = & & 2\text{ Re } \Bigg[\bra{\chi_0(t_f)} \nabla U_{\mu,\vec{\lambda}_0} (t_f,0) \ket{\chi_0(0)}\nonumber\\
& & \bra{\chi_0(t_f)} U_{\mu,\vec{\lambda}_0} (t_f,0) \ket{\chi_0(0)}^*\Bigg],
\end{eqnarray}
can be expand in $\mu$
\begin{eqnarray}
\nabla F (\mu, \lambda_0)
= \sum_{n=0}^\infty \mu^n \vec{\cF}^{(n)}.
\label{app_gradF}
\end{eqnarray}
Note that $\nabla$ is the gradient with respect to $\vec\lambda$.

If we define
\begin{eqnarray}
\vec{v}_j = \bra{\chi_0 (t_f)} \nabla U_{\vec\lambda_0}^{(j)} (t_f,0) \ket{\chi_0 (0)},
\end{eqnarray}
then we can express the coefficients in Eq. \eqref{app_gradF} as
\begin{eqnarray}
\vec{\cF}^{(n)}=2 \mbox{ Re } \left( \sum_{k=0}^n  \vec{v}_{n-k} u_k^* \right).
\end{eqnarray}

% --- v0 and cF0 ---
It will be useful to define a generalisation of $\vec v_0$ namely
\begin{eqnarray}
\vec{W}_{n,m} (t_2, t_1) \equiv \braXket{\chi_n (t_2)}{\nabla U_{\vec\lambda_0}^{(0)} (t_2,t_1)}{\chi_m (t_1)}.
\end{eqnarray}
Using time-dependent perturbation theory (similar to the previous subsection) and $\Hamil^{(0)} (\vec\lambda) \approx \Hamil^{(0)} (\vec\lambda_0) + (\vec\lambda-\vec\lambda_0) \cdot \nabla \Hamil^{(0)} (\vec\lambda_0)$, simplifies this to
\begin{eqnarray}
\vec{W}_{n,m} (t_2,t_1) &=& -\frac{i}{\hbar} \int_{t_1}^{t_2} dt \, \vec{\beta}^{(0)}_{n,m} (t).
\end{eqnarray}
Especially $\vec{v}_0 = \vec{W}_{0,0} (t_f,0)$.
Since $\vec{W}_{0,0}$ is purely imaginary, it follows that
$\vec{\cF}^{(0)} = \vec{0}$.

% --- v1 and cF1
\begin{widetext}
From Eq. \eqref{U_first}, we get 
\begin{eqnarray}
\vec{v}_1 = - \frac{i}{\hbar} \int_{0}^{t_f} dt \bra{\chi_0 (t_f)} \nabla \left. \left[ U_{\vec\lambda}^{(0)} (t_f,t)
\Hamil^{(1)} (\vec\lambda;t) U_{\vec\lambda}^{(0)} (t,0) \right] \right\rvert_{\vec\lambda=\vec\lambda_0} \ket {\chi_0 (0)}.
\end{eqnarray}
Using the product rule and inserting identities $\unit = \sum_n \ketbra{\chi_n(t)}{\chi_n(t)}$
we arrive at
\begin{eqnarray}
\vec{v}_1 &=& - \frac{i}{\hbar} \int_{0}^{t_f} dt \left\{ \sum_n \left[ \vec{W}_{0,n} (t_f, t) \alpha^{(1)}_{n,0} (t)
+ \alpha^{(1)}_{0,n} (t) \vec{W}_{n,0} (t,0)\right] + \vec{\beta}^{(1)}_{0,0} (t) \right\} \nonumber\\
&=& -\frac{1}{\hbar^2} \sum_n \int_0^{t_f} dt
\left[ \int_t^{t_f} ds \, \vec{\beta}^{(0)}_{0,n} (s) \alpha^{(1)}_{n,0} (t)  + \int_0^{t} ds \, \alpha^{(1)}_{0,n} (t) \vec{\beta}^{(0)}_{n,0} (s)\right] - \frac{i}{\hbar} \int_0^{t_f} dt \, \vec{\beta}^{(1)}_{0,0} (t) .
\end{eqnarray}
%\end{widetext}
In order to find $\vec{\cF}^{(1)}$, we first calculate
\begin{eqnarray}
2 \text{ Re } (\vec{v}_1) &=&
-\frac{2}{\hbar^2} \sum_n \text{Re } \left[\int_0^{t_f} dt \, \alpha^{(1)}_{n,0} (t)
\int_0^{t_f} ds\, {\vec{\beta}^{(0)}_{n,0} (s)}^* \right],
\end{eqnarray}
and
\begin{eqnarray}
2 \text{Re } (\vec{v}_0 u_1^*) &=&
\frac{2}{\hbar^2} \text{Re }\left[\int_0^{t_f} dt \alpha^{(1)}_{0,0} (t)
\int_0^{t_f} ds\, {\vec{\beta}^{(0)}_{0,0} (s)}^* \right].
\end{eqnarray}
By combining these two results gives
\begin{eqnarray}
\vec{\cF}^{(1)} = 
-\frac{2}{\hbar^2} \sum_{n=1}^\infty \text{Re } \left[\int_0^{t_f} dt \, \alpha^{(1)}_{n,0} (t)
\int_0^{t_f} ds\, {\vec{\beta}^{(0)}_{n,0} (s)}^* \right].
\end{eqnarray}

% --- v2 and cF2
Using Eq. \eqref{U_second} and a similar calculation as above, we arrive at
%\begin{widetext}
\begin{eqnarray}
&&\vec{v}_2 \nonumber \\&=& \frac{i}{\hbar^3} \int_{0}^{t_f} dt \int_0^t ds\, \sum_{n,m} \left[ \int_t^{t_f}\! du\, \vec{\beta}^{(0)}_{0,n} (u) \alpha^{(1)}_{n,m} (t) \alpha^{(1)}_{m,0} (s) + \int_s^{t}\! du\, \alpha^{(1)}_{0,n} (t) \vec{\beta}^{(0)}_{n,m} (u) \alpha^{(1)}_{m,0} (s)+ \int_0^{s}\! du\, \alpha^{(1)}_{0,n} (t) \alpha^{(1)}_{n,m} (s) \vec{\beta}^{(0)}_{m,0} (u)\right]\nonumber\\
&-& \frac{1}{\hbar^2} \int_{0}^{t_f} dt \int_0^t ds\, \left\{\sum_n \left[ \vec{\beta}^{(1)}_{0,n} (t) \alpha^{(1)}_{n,0} (s)
+ \alpha^{(1)}_{0,n} (t) \vec{\beta}^{(1)}_{n,0} (s) \right]
\right\}\ \nonumber\\
&-&
\frac{1}{\hbar^2} \sum_n \int_0^{t_f} dt
\left[ \int_t^{t_f} ds \vec{\beta}^{(0)}_{0,n} (s) \alpha^{(2)}_{n,0} (t)
+ \int_0^{t} ds \alpha^{(2)}_{0,n} (t) \vec{\beta}^{(0)}_{n,0} (s)\right]
- \frac{i}{\hbar} \int_0^{t_f} dt \, \vec{\beta}^{(2)}_{0,0} (t) .
\end{eqnarray}
Similarly to the calculation of $\vec{\cF}^{(2)}$, we neglect contributions involving three integrals over time which results in
\begin{eqnarray}
\vec{\cF}^{(2)} \approx - \frac{2}{\hbar^2} \sum_{n=1}^\infty \mbox{ Re } \left[
\left(\int_0^{t_f} dt \, \vec{\beta}^{(1)}_{n,0}(t)\right)^* \left(\int_0^{t_f} dt \, \alpha^{(1)}_{n,0}(t)\right)+ \left(\int_0^{t_f} dt \, \vec{\beta}^{(0)}_{n,0}(t)\right)^* \left(\int_0^{t_f} dt \, \alpha^{(2)}_{n,0}(t)\right)\right].
\end{eqnarray}
\end{widetext}
\end{appendix}

\bibliography{eSTA_paper}{}
\bibliographystyle{apsrev4-1}

\end{document}